\documentclass[apjl]{emulateapj}

\usepackage{hyperref}
\usepackage{amsmath}
\usepackage{amssymb}
\usepackage{graphicx}

\newcommand{\etaearth}{\eta_\oplus}
\newcommand{\Rpeak}{R_\mathrm{peak}}
\newcommand{\REarth}{R_\oplus}
\newcommand{\RSun}{R_\odot}
\newcommand{\RStar}{R_\mathrm{star}}
\newcommand{\MStar}{M_\mathrm{star}}
\newcommand{\MSun}{M_\odot}
\newcommand{\Rpl}{\Lambda_\mathrm{pl}}
\newcommand{\Npl}{N_\mathrm{pl}}

\newcommand{\Rbg}{\Lambda_\mathrm{bg}}
\newcommand{\Nbg}{N_\mathrm{bg}}

\newcommand{\rhomin}{\rho_\mathrm{min}}
\newcommand{\rhomax}{\rho_\mathrm{max}}

\newcommand{\earange}{3.9_{-1.6}^{+2.2}\%}
\newcommand{\rplrange}{3.83_{-0.62}^{+0.76}}
\newcommand{\rpeakrange}{1.25_{-0.17}^{+0.16}}
\newcommand{\corrcoeffrange}{0.334_{-0.053}^{+0.052}}
\newcommand{\fposrange}{7.8_{-1.3}^{+1.4}\%}
\newcommand{\ppeakrange}{0.075_{-0.006}^{+0.007}}
\newcommand{\rhominrange}{5.46^{+0.18}_{-0.18}}
\newcommand{\rhomaxrange}{18.8^{+1.9}_{-1.9}}

\begin{document}

\title{The Occurrence of Earth-Like Planets Around Other Stars}
\shorttitle{Earth-Like Planets Around Other Stars}
\author{Will M. Farr, Ilya Mandel, Chris Aldridge \and Kirsty Stroud}
\affil{School of Physics and Astronomy\\University of
  Birmingham\\Birmingham\\B15 2TT\\United Kingdom}
\email{w.farr@bham.ac.uk}
\email{ilyamandel@chgk.info}
\email{cxa064@bham.ac.uk}
\email{kls081@bham.ac.uk}

\begin{abstract}
  The quantity $\etaearth$, the number density of planets per star per
  logarithmic planetary radius per logarithmic orbital period at one
  Earth radius and one year period, describes the occurrence of
  Earth-like extrasolar planets.  Here we present a measurement of
  $\etaearth$ from a parameterised forward model of the (correlated)
  period-radius distribution and the observational selection function
  in the most recent (Q17) data release from the Kepler satellite.  We
  find $\etaearth = \earange$ (90\% CL).  We conclude that each star
  hosts $\rplrange$ planets with $P \lesssim 3 \mathrm{yr}$ and $R
  \gtrsim 0.2 \REarth$.  Our empirical model for false-positive
  contamination is consistent with the dominant source being
  background eclipsing binary stars.  The distribution of planets we
  infer is consistent with a highly-stochastic planet formation
  process producing many correlated, fractional changes in planet
  sizes and orbits.
\end{abstract}


\maketitle

\section{Introduction}

The quantity $\etaearth$, the number density of planets per star per
logarithmic planetary radius per logarithmic orbital period at one
Earth radius and one year period, describes the occurrence of
Earth-like extrasolar planets.  Measurement of $\etaearth$ is
complicated by the difficulty of detecting Earth-like planets in
Earth-like orbits about Sun-like stars.  Here we present a measurement
of $\etaearth$ from a parameterised forward model of the (correlated)
period-radius distribution and the observational selection function in
the most recent (Q17) data release from the Kepler satellite
\citep{Borucki2010,Borucki2011,Batalha2013}.  Our data set comprises
181,568 systems observed under the Kepler exoplanet observing program
(mostly G-type stars on the main sequence \citep{Batalha2010}),
producing 2598 planetary candidates.  We parameterise the distribution
of planetary periods and radii using a single, correlated Gaussian
component; treat selection effects using a parameterised transit
detection probability based on the measured noise level and stellar
properties in the Kepler catalog; and include an
empirically-parameterised, independent component in the period-radius
distribution to represent false-positive planet detections.  Using our
model we can simultaneously estimate $\etaearth$, place constraints on
the planet period-radius distribution function, and determine the
degree of contamination by false-positive candidate identifications.
We find $\etaearth = \earange$ (90\% CL).  We conclude that each star
hosts $\rplrange$ planets with $P \lesssim 3 \mathrm{yr}$ and $R
\gtrsim 0.2 \REarth$, that the peak of the planet radius distribution
lies at $\Rpeak = \rpeakrange \REarth$, and that $\ln P$ and $\ln R$
are correlated with correlation coefficient $r = \corrcoeffrange$ (all
90\% CL).  Our empirical model for false-positive contamination is
consistent with the dominant source being background eclipsing binary
stars \citep{Fressin2013}, with $\fposrange$ (90\% CL) of the
candidates being false-positives.  The distribution of planets we
infer is consistent with a highly-stochastic planet formation process
producing many correlated, fractional changes in planet sizes and
orbits.  Our approach of determining both the intrinsic distribution
of objects and selection effects empirically from survey data is
generally applicable.

The Kepler satellite detects planets by observing a decrement in the
photometric intensity of a planet's host star as the planet transits
between the telescope and the star.  The Q17 data release describes
2598 ``candidate'' planetary transit signals identified by the Kepler
team from observations of stars in the ``EX'' observing program (which
are primarily G-type main-sequence stars similar to our own Sun
\citep{Batalha2010}), giving the inferred planetary period and radius
for each.  The fractional depth of a planetary transit signal depends
only on the radii of the planet and its host star.  The signal to
noise ratio of a series of transits about a particular star in the
Kepler satellite scales with planetary period and radius
as \citep{Chatterjee2012}
\begin{equation}
  \rho = \rho_0 \left( \frac{R}{\REarth} \right)^2 \left(
  \frac{P}{1\,\mathrm{yr}} \right)^{-1/3},
\end{equation}
where $\rho_0$ is the signal to noise ratio of a Earth-radius planet
in a one-year orbit about that star, which depends on the number of
quarters of observation of that star, the stellar radius and mass, and
the intrinsic variability of the stellar intensity
\citep{Christiansen2012}.  In our analysis, we obtain these quantities
from the Kepler Input Catalog \citep{Batalha2010,Brown2011} and the
MAST Kepler archive\footnote{\url{http://archive.stsci.edu/kepler/}}.

\section{Model}

\begin{figure}
  \includegraphics[width=\columnwidth]{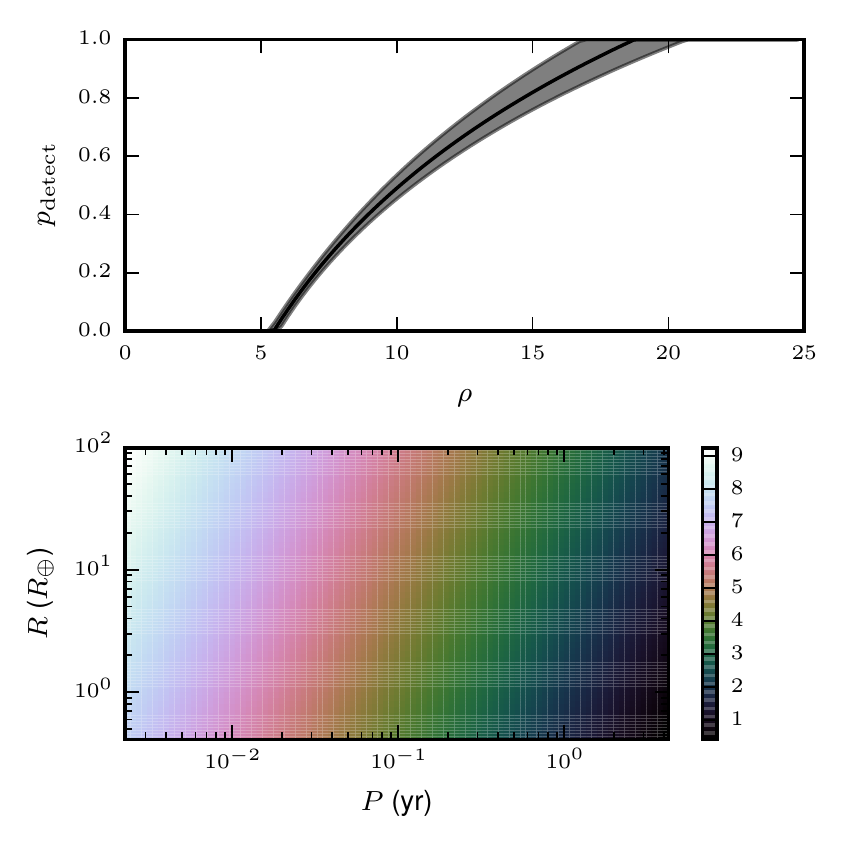}
  \caption{\label{fig:det-bg} \textbf{Inferred detection probability
      and density of background contamination.} (Top) The inferred
    detection probability versus signal-to-noise ratio (see
    Eq.\ \eqref{eq:pdetect}) from our parameterised model of selection
    effects.  The solid line is the posterior median detection
    probability and the shading gives the 90\% credible posterior
    interval.  Our inferred detection probability is in rough
    agreement with the measurements of detection efficiency in
    \citet{Borucki2011,Batalha2013}.  (Bottom) The number density of
    false-positive candidate signals, $d\Nbg/d\ln P \, d\ln R$ (see
    Eq.\ \eqref{eq:background-rate}).  The density is highest at small
    candidate period and large radius, consistent with the dominant
    source of contamination being background eclipsing
    binaries \citep{Fressin2013}.  Overall, our model finds
    $\fposrange$ of the candidates are false-positive background
    signals, consistent with the analysis in \citet{Fressin2013}.}
\end{figure}

To a good approximation (see Fig.\ \ref{fig:selection} below), the
detectability of a series of planetary transits in the Kepler data set
is a function of the signal to noise ratio of the series.  Because the
detectability of planet transits depends on both period and radius, it
is important to consider the joint (i.e., two-dimensional)
distribution of these quantities in the data
\citep{Tabachnik2002,Youdin2011}.  We model the detection probability
of a transit as a function that rises linearly in the log of the
signal to noise ratio from zero at a threshold signal to noise to one
at a larger signal to noise:
\begin{equation}
  \label{eq:pdetect}
  p_\mathrm{detect} = \begin{cases}
    0 & \rho < \rhomin \\
    \frac{\log \rho - \log \rhomin}{\log \rhomax - \log \rhomin} &
    \rhomin < \rho < \rhomax \\
    1 & \rhomax < \rho
  \end{cases},
\end{equation}
where $\rhomin$ and $\rhomax$ are parameters of our model.  We find
$\rhomin = \rhominrange$ and $\rhomax = \rhomaxrange$ (90\% CL), in
rough agreement with \citet{Borucki2011,Batalha2013}.  A plot of our
inferred detection probability appears in Fig.\ \ref{fig:det-bg}

The probability that a planet's orbital plane will align with the
line-of-sight to Earth and thereby produce a transit signal is
\begin{equation}
  \label{eq:ptransit}
  p_\mathrm{transit} = 0.0016\, \frac{\RStar}{\RSun}
  \left(\frac{\MStar}{\MSun}\right)^{-1/3} \left(\frac{P}{1\,\mathrm{yr}}\right)^{-2/3}.
\end{equation}
Putting Eq.\ \ref{eq:pdetect} and \ref{eq:ptransit} together, the
probability that Kepler will detect a planet of radius $R$ orbiting
its host star at period $P$ is 
\begin{equation}
  \label{eq:pselect}
  p_\mathrm{select} = p_\mathrm{transit} p_\mathrm{detect}.
\end{equation}

A correlated log-normal distribution of planets in period and radius
would be a natural outcome of a stochastic planet formation process
that produced many correlated, fractional changes in planet sizes and
orbits.  As we shall see (Figure\ \ref{fig:selection}), this simple
model combined with the aforementioned selection function provides a
good fit to the Kepler candidate distribution.  In our model, observed
planets populate the candidate $P$-$R$ plane with number density
\begin{multline}
  \label{eq:foreground-rate}
  \frac{dN_\mathrm{obs}}{d\ln P\, d\ln R} = \left[ \sum_\mathrm{stars}
    p_\mathrm{select}(P, R) \right] \\ \times \Rpl N\left[ \mu, \Sigma
    \right]\left( \ln P, \ln R \right),
\end{multline}
where $\Rpl$, $\mu$, and $\Sigma$ are parameters of our model, with
$\Rpl$ the average number of planets per star, $\mu = \left[ \mu_P,
  \mu_R \right]$ the mean of $\ln P$ and $\ln R$, and $\Sigma = \left[
  \left[ \Sigma_{PP}, \Sigma_{PR} \right], \left[ \Sigma_{PR},
    \Sigma_{RR} \right]\right]$ the covariance matrix of $\ln P$ and
$\ln R$; $N\left[ \mu, \Sigma \right](x,y)$ is the normal
distribution.  Our model assumes that planets appear around their host
stars in a Poisson process; this is almost certainly wrong in detail
\citep{Weissbein2012}, but nevertheless provides a good fit to the
observed data (see Figure \ref{fig:selection}).

In addition to true planetary signals, we model a false-positive
background of planet candidates empirically, assuming they populate
the candidate $P$-$R$ plane with a number density that has a linear
gradient across a rectangular region in the $\ln P$-$\ln R$ plane:
\begin{multline}
  \label{eq:background-rate}
  \frac{d\Nbg}{d \ln P \, d \ln R} = \frac{\Nbg}{\Delta \ln P \Delta
    \ln R} \\ \times \left( 1 + \vec{\gamma} \cdot \left[ \ln P - \ln P_\mathrm{mid} , \ln
    R - \ln R_\mathrm{mid} \right] \right),
\end{multline}
where $\Delta \ln P = \ln P_\mathrm{max} - \ln P_\mathrm{min}$, $\ln
P_\mathrm{mid} = 1/2\left(\ln P_\mathrm{max} - \ln P_\mathrm{min}
\right)$, $\Delta \ln R = \ln R_\mathrm{max} - \ln R_\mathrm{min}$,
$\ln R_\mathrm{mid} = 1/2\left(\ln R_\mathrm{max} - \ln R_\mathrm{min}
\right)$.  $\Nbg$, the expected number of background false-positive
events; $P_\mathrm{max}$, $P_\mathrm{min}$, $R_\mathrm{max}$, and
$R_\mathrm{min}$, the boundaries in the $P$-$R$ plane within which
background events appear; and $\gamma$, the gradient in the number
density of background events, are parameters of our model.  This is a
purely empirical model for the background contamination, but is
reasonable if the chief contaminant is background eclipsing binaries
\citep{Fressin2013,Duquennoy1991}.  The posterior on the background
number density in the $P$-$R$ plane appears in
Figure\ \ref{fig:det-bg}.

Unlike \citet{Foreman-Mackey2014}, we do not attempt to model the
observational uncertainties in the estimated periods and radii from
the Kepler candidate data set.  In spite of several candidates with
very large uncertainties in measured parameters, we have found that
our fit is essentially unchanged when applied to synthetic
observations with periods and radii re-drawn from the range of
observational uncertainties quoted in the Q17 data release.

The likelihood of the observed periods and radii under our model is an
inhomogeneous Poisson likelihood \citep{Farr2013,Youdin2011} with a
rate that is the sum of Eq.\ \eqref{eq:foreground-rate} and
Eq.\ \eqref{eq:background-rate}.  We impose priors on our 15 model
parameters as follows: for the planet occurrence rate $\Rpl$ and
(implicitly) the parameters describing selection effects, we impose a
$1/\sqrt{\Npl}$ prior; for the background rate $\Rbg$ we impose a
$1/\sqrt{\Rbg}$ prior; for the selection model parameters $\rhomin$
and $\rhomax$ we impose a log-normal prior with unit width at signal
to noise ratios of 3 and 11, respectively; in all other parameters we
impose a flat (i.e., constant-density) prior.  The product of
likelihood and prior gives a Bayesian posterior density function on
the fifteen-dimensional parameter space of our model.  We sample from
this function using the \texttt{emcee} sampler
\citep{Foreman-Mackey2013}.  The posterior describes simultaneously
the intrinsic distribution and number of exoplanets, the amount and
distribution of the contaminating false-positive events in the
candidate data set, and the selection function of the instrument for
true planetary transit events.

\section{Conclusion}

\begin{figure}
  \includegraphics[width=\columnwidth]{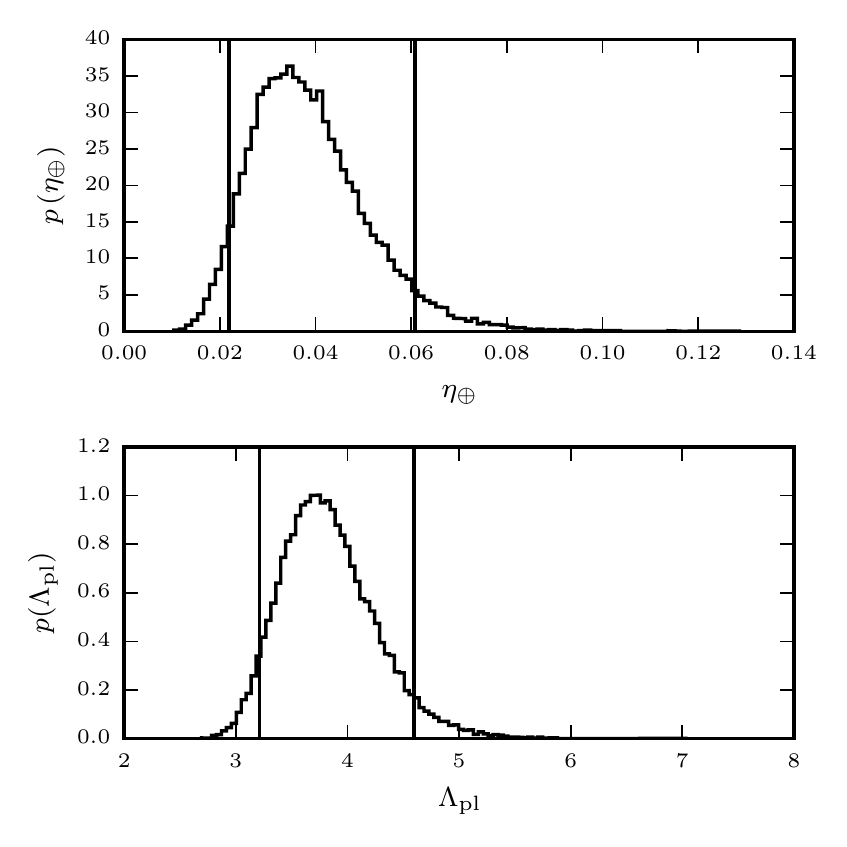}
  \caption{\label{fig:eta-earth} \textbf{Posteriors on $\etaearth$ and
      $\Rpl$ accounting for selection effects and false-positive
      detections.}  (Top) The inferred posterior density on $\etaearth
    = dN/d\ln P \, d\ln R \left( 1\, \textnormal{yr}, R_\oplus
    \right)$.  Vertical lines indicate the 90\% credible range.  We
    find $\etaearth = \earange$.  (Bottom) The inferred posterior on
    $\Rpl$, the number of planets per star with $P \lesssim 3
    \mathrm{yr}$ and $R \gtrsim 0.2 \REarth$.  Vertical lines indicate
    the 90\% credible range.  We find $\Rpl = \rplrange$.}
\end{figure}

The main result of this paper, the posterior distribution for
$\etaearth$, the number density of Earth-like planets, marginalised
over all other parameters in our model (i.e., incorporating our
uncertainty about contamination, selection effects, intrinsic
distribution of planets, etc) appears in Fig.\ \ref{fig:eta-earth}.
Recall that
\begin{multline}
  \etaearth = \left. \frac{dN}{d \ln P \ln R} \right|_{R = \REarth, P
    = 1\,\mathrm{yr}} \\ = \Rpl N\left[ \mu, \Sigma \right]\left( \ln
  1\,\mathrm{yr}, \ln \REarth \right),
\end{multline}
which is roughly the number of planets per star with periods and radii
within a factor of $\sqrt{e}$ of Earth's.  We find $\etaearth =
\earange$ (90\% CL).  Our model also gives an estimate of the number
of planets of any radius and period per star; the posterior for this
quantity, marginalised over all other parameters also appears in
Fig.\ \ref{fig:eta-earth}.  We find $\Rpl = \rplrange$ (90\% CL).

\begin{figure}
  \includegraphics[width=\columnwidth]{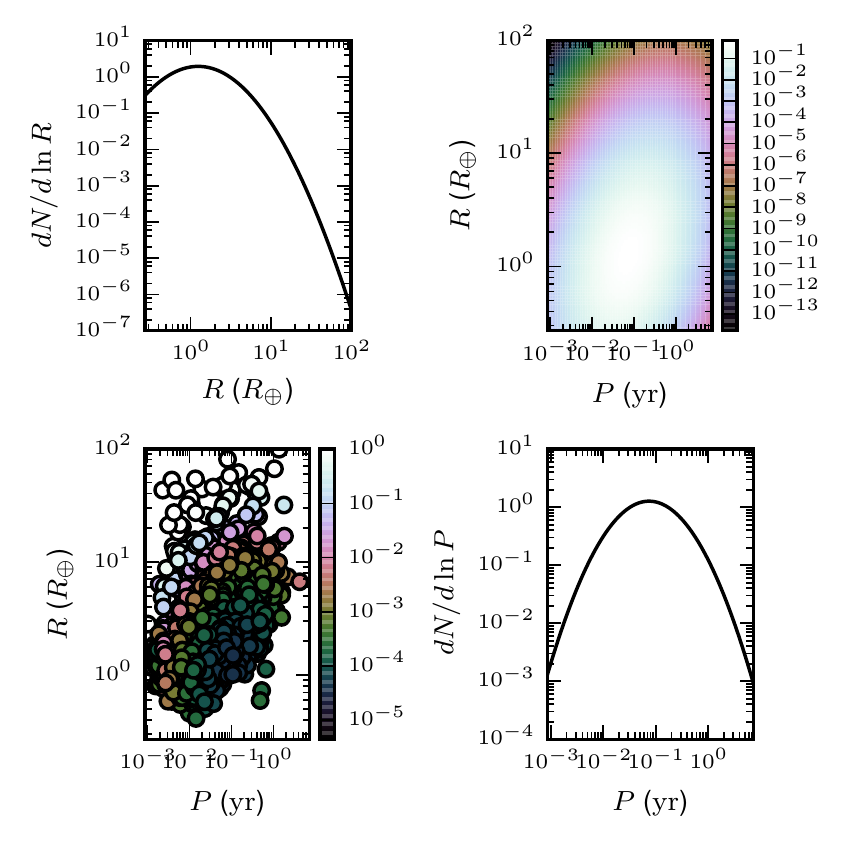}
  \caption{\label{fig:foreground-dist} \textbf{The inferred planet
      period--radius distribution accounting for selection effects and
      false-positives.}  (Upper Left) The planet number density per
    logarithmic planet radius.  The density peaks at $\Rpeak =
    \rpeakrange \REarth$ (90\% CL).  (Upper Right) The planet number
    density in the period--radius plane.  The inferred correlation
    coefficient between $\ln P$ and $\ln R$ is $r = \corrcoeffrange$.
    (Lower Left) Scatter plot of the radius and period of the Kepler
    planet candidates.  Color indicates the posterior false-positive
    probability for each candidate.  Overall, the model prefers a
    false-positive rate of $\fposrange$ (90\% CL).  The primary
    contaminant is probably background eclipsing binaries; our
    contamination rate is consistent with previous work
    \citep{Fressin2013}. (Lower Right) The planet number density per
    logarithmic planet period.  The density peaks at $P = \ppeakrange
    \mathrm{yr}$ (90\% CL). }
\end{figure}

Our model allows us to produce a posterior on the distribution of
planets in the period-radius plane, and the probability that any given
planetary candidate is a planet instead of a background contaminant;
these posteriors appear in Fig. \ref{fig:foreground-dist}.  Our model
finds that the false-positive rate in the candidate data set is
$\fposrange$ (90\% CL), consistent with previous work
\citep{Fressin2013} estimating the contamination in the Kepler
candidate set.  Our model has the peak of the planet period-radius
distribution at $\Rpeak = \rpeakrange \REarth$, $P_\mathrm{peak} =
\ppeakrange \mathrm{yr}$, and the distribution of planetary radii and
periods is correlated, with correlation coefficient $r =
\corrcoeffrange$ (all at 90\% CL).

\begin{figure}
  \includegraphics[width=\columnwidth]{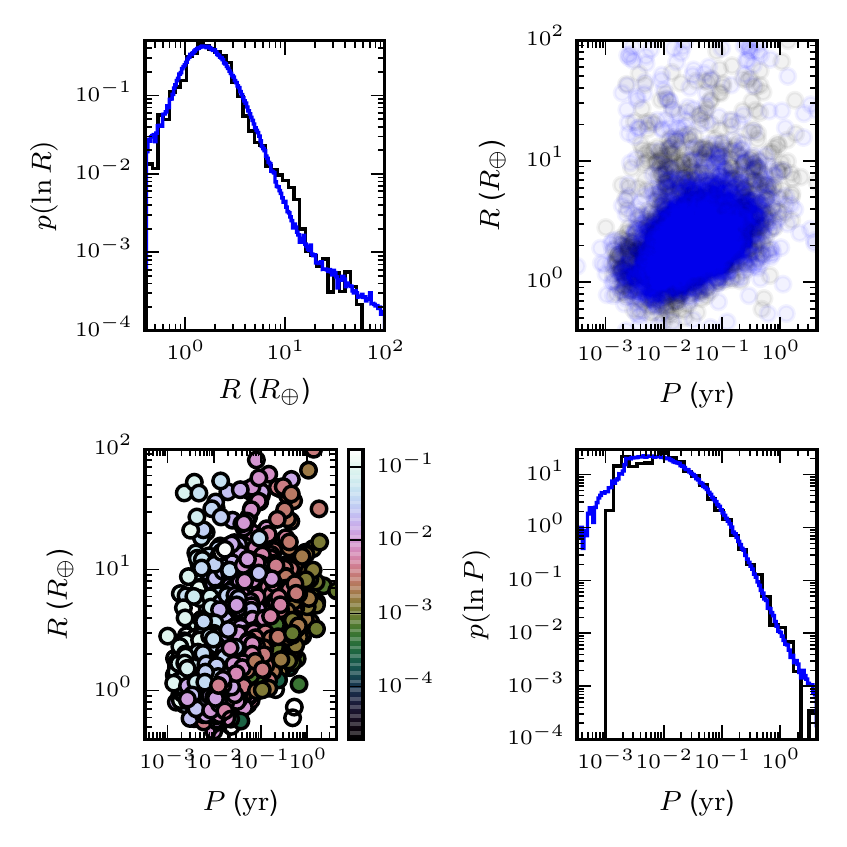}
  \caption{\label{fig:selection} \textbf{Comparison of synthetic data
      sets produced from the forward model incorporating selection
      effects with observed candidates.}  (Upper Left) The observed
    (black curve) and synthetic (blue curve) normalised candidate
    density per logarithmic radius.  Except for a discrepancy at $R
    \simeq 10 \REarth$---associated with hot Jupiters, a distinct
    planetary population \citep{Albrecht2012,Naoz2012}---the model
    produces a good fit to the observed candidates over the range of
    reported radii.  Note particularly the tail at large radii that
    comes from background contaminants in both observed and synthetic
    data.  (Upper Right) Scatter plot of the observed candidates
    (black circles) and a posterior-averaged draw of observed
    candidates from the model (blue circles).  (Lower Left) Scatter
    plot of the observed candidates.  Colors indicate the
    posterior-averaged selection probability for each planet about its
    host star (see Eq.\ \eqref{eq:pselect}).  (Lower Right) The
    observed (black curve) and synthetic (blue curve) normalised
    candidate density per logarithmic period.  Except for the
    aforementioned hot Jupiter peak at $P \simeq 1 \mathrm{day}$ the
    model produces a good fit to the observed candidates over the
    range of reported periods.}
\end{figure}

Our model predicts a distribution for future observed data consistent
with the already-observed candidate set.  These predictions can be
used to perform graphical and posterior-predictive model checking
\citep{Gelman2013}.  Fig.\ \ref{fig:selection} compares the
predictions of our model for observed periods and radii (incorporating
both planetary transits and background events) with the candidate set.
This is a particularly stringent test of our parameterised selection
model since the observed periods and radii are strongly influenced by
the selection function of the Kepler telescope and pipeline.  Except
for the known sub-population of hot Jupiters
\citep{Albrecht2012,Naoz2012}, our model provides a very good fit to
the observed data.  That a simple log-normal distribution in period
and radius fits the observed distribution of planets well may indicate
that planet formation is a stochastic process with many small,
correlated, and multiplicative influences on planet period and radius
resulting, from the central limit theorem, in a log-normal
distribution in these parameters.

Previous estimates
\citep{Catanzarite2011,Traub2012,Dong2013,Petigura2013,Foreman-Mackey2014}
place $1\% \lesssim \etaearth \lesssim 34\%$.  These works dealt with
the problem of selection effects in the sample by either analysing a
region of the period-radius parameter space where observations are
complete and extrapolating to $R = \REarth$ and $P = 1 \mathrm{yr}$
\citep{Catanzarite2011,Traub2012}, applying a binned analysis
incorporating survey incompleteness in the period-radius plane
\citep{Dong2013,Petigura2013} or analysing the results of a customised
planet detection pipeline on a subset of the Kepler observations
\citep{Petigura2013,Foreman-Mackey2014}.  The methods and analysed
data sets of \citet{Petigura2013,Foreman-Mackey2014} are most
comparable to ours.  These studies used the same data set, produced
\citep{Petigura2013} from a subset of the available Kepler data and a
customised pipeline to search for transit signals.  They both
accounted for selection effects by measuring the recoverability of
synthetic transit signals injected into their data, in contrast to our
approach of empirically determining them from the observed data.
Neither study attempted to account for contamination from
falsely-identified candidate transit events, controlling this instead
through careful choice of threshold.  Both studies used a more
flexible model for the intrinsic distribution of planets than ours.
Our result for $\etaearth$ is consistent with, but more precise than,
\citet{Foreman-Mackey2014} and (somewhat) inconsistent with
\citet{Petigura2013}.

\acknowledgements

The code implementing this analysis is available under an open-source
``MIT'' license at \url{https://github.com/farr/kepler-selection}.
This work was supported by the Science and Technology Facilities
Council.  Computations in this work were performed on the University
of Birmingham's BlueBEAR cluster.  Some of the data presented in this
paper were obtained from the Mikulski Archive for Space Telescopes
(MAST). STScI is operated by the Association of Universities for
Research in Astronomy, Inc., under NASA contract NAS5-26555. Support
for MAST for non-HST data is provided by the NASA Office of Space
Science via grant NNX13AC07G and by other grants and contracts.  This
paper includes data collected by the Kepler mission. Funding for the
Kepler mission is provided by the NASA Science Mission directorate.

\bibliography{kepler-selection}

\end{document}